# Autonomous Observations in Antarctica with AMICA

**Gianluca Di Rico[1], Maurizio Ragni[1], Mauro Dolci[1], Oscar Straniero[1], Angelo Valentini[1], Gaetano Valentini[1], Amico Di Cianno[1], Croce Giuliani[1], Demetrio Magrin[2], Carlotta Bonoli[2], Favio Bortoletto[2], Maurizio D'Alessandro[2], Leonardo Corcione[3] and Alberto Riva[3]**

[1]*INAF - Osservatorio Astronomico di Teramo, Teramo, ITALY*
[2]*INAF - Osservatorio Astronomico di Padova, Padova, ITALY*
[3]*INAF – Osservatorio Astronomico di Torino, Pino Torinese (TO), ITALY*

The Antarctic Multiband Infrared Camera (AMICA) is a double channel camera operating in the 2-28 μm infrared domain (KLMNQ bands) that will allow to characterize and exploit the exceptional advantages for Astronomy, expected from Dome C in Antarctica. The development of the camera control system is at its final stage. After the investigation of appropriate solutions against the critical environment, a reliable instrumentation has been developed. It is currently being integrated and tested to ensure the correct execution of automatic operations. Once it will be mounted on the International Robotic Antarctic Infrared Telescope (IRAIT), AMICA and its equipment will contribute to the accomplishment of a fully autonomous observatory.

## 1. Introduction

In the recent years, great attention has been paid to the new possibilities opened from the exploitation of Antarctica for astronomical observations. A large number of advantages come out from the peculiar characteristics of this remote land and, at the same time, new stations are rising in the inner region of the continent, known as Antarctic Plateau. Among them *Concordia*, one of the most advanced and organized permanent bases, built at Dome C within a French-Italian collaboration.

Taking advantage of the presence of this station, an ambitious goal is going to be achieved, consisting in the installation of a fully autonomous observatory, constituted by IRAIT [1] and the scientific equipment of AMICA [2]. IRAIT is a 0.8 m, F/22 Cassegrain with two Nasmyth foci, built within an Italian-Spanish collaboration among the University of Perugia, the University of Granada (DFTC) and the Institut d'Estudis Espacials de Barcelona (IEEC). Its mechanics has been mounted during the 2008-09 summer summer campaign while the assembly of the remaining components, the mirrors alignment and the integration of the camera system will be accomplished strating from the 2009-10 summer.

Main scientific tasks of AMICA are both the characterization of the Dome C sky for infrared astronomy and the observation of a large variety of astrophysical objects. Among them the AGB and post-AGB stars, the star forming regions in our and nearby galaxies, but also RR-lyrae, nearby brown dwarf, heavly-obscured supernovae and Solar System bodies. Finally, Survey mode observations will be performed for interesting regions of the southern sky (e.g. the Large and the Small Magellanic Clouds).

The robotization of the whole system is a necessary condition due to the extreme climate conditions, as a result of which human activities are reduced and essentially stopped during the Antarctic winter (mainly because of the unavailability of efficient communication means and the total absence of personnel and supplies transport). On the other hand, the instrumentation needs reliable solutions to deal with such peculiar environment, to avoid the damage of the components and to minimize any safety risk.

## 2. The Antarctic Scenario

In this section, we give a brief overview of pros and cons of the Antarctic environment from both a scientific and technological point of view. These considerations have led to the development of AMICA and to the adoption of the solutions discussed afterwards.

### 2.1. Sky Properties

Essential requirements for observations at infrared wavelengths are the low sky emission and the high atmospheric transmission. As reported in literature, first estimations of these properties in Antarctica and in particular in the inner Plateau are highly promising [3,4]. The extremely low temperatures (with an annual mean of about -55°C) reduce the thermal emission from the sky and prevent the atmosphere from containing high amount of water vapor. Since the most absorption is dominated by this component, near infrared bands (2-5 μm) are wider than those in a temperate site and new windows open up in the mid-infrared (beyond 15 μm), which cannot be accessed elsewhere [5]. Other important advantages come out from the low level of aerosols and dust, the high stability of the atmosphere, the high percentage of cloud-free time and the elevation of the site (3250 m). For these reasons, a substantial increment of the resulting sensitivity is expected, obtaining, for a given primary mirror size, the same performance of a several times larger telescope, with lower construction and management costs.

## 2.2. Polar Condition

Further important characteristics have to be considered about the location of Dome C. Its near-Pole latitude (75°S) allows to perform longer and uninterrupted observations in a maximum field of circumpolar sources (particularly interesting for the presence of peculiar regions such as the Magellanic Clouds and the Galaxy Center), even at zenithal angles normally unfavorable for temperate sites. Finally, thanks to the excellent IR properties of the sky, daily duty-cycle of 100% is probably achievable for wavelengths beyond 4μm, where observations can be carried out even with sunlight.

## 2.3. The Extreme Environment

Unfortunately, the site location and the environmental conditions that are responsible for the great advantages underlined above, make it very difficult to operate conventional instrumentation. Main troubles rise out from the temperature and pressure values.

*Low Temperature*

The first problem to face concerns the low temperatures (down to -80°C during wintertime). Despite they provide an efficient passive cooling (dramatically reducing the instrumental emission that at infrared wavelengths could overcome the sky flux), the electronic and mechanical systems are dangerously exposed to the risk of damage and malfunction. The investigation of suitable solutions for components that have to inevitably work outdoors, and the insulation of devices with limited operating conditions, are very critical aspects. In fact, they require the development of custom elements accurately tested in climatic chamber, to simulate the environment in which they will be operated and stored.

*Low Pressure*

Because of the low pressure (~ 640 mbar at Dome C), the efficiency of the conduction and the convection of the air is significantly reduced, hence increasing the possible overheating of insulated electronics. This problem has been frequently reported during experiments performed in past Antarctic campaigns [6]. For this reason a thermal study of dissipating elements and a careful distribution of hot spots inside the cabinets are fundamental prerogatives for an efficient active conditioning system.

Temperature and pressure are also responsible of the limited human activities, in particular during winter, when the difficulty to operate outdoors becomes prohibitive.

## 2.4. Logistic Facilities

The presence of the Concordia base ensures the availability of a large number of facilities for the permanence of the personnel, the visiting researchers and the hosted scientific experiments. However some relevant restrictions need to be considered:

- the total amount of electrical supply that can be provided by the station generators is 200 kW (shared among all scientific and logistic needs);
- satellite links are normally used for fax and voice connections. Data transfer is very limited in time and speed (~ 64 kbps) and used mainly for text emails. Some improvements are under study to provide a broad-band data connection.
- the site can be reached by plane from the Italian "Mario Zucchelli" coastal station after a 5 hours flight, or by ground from the French coastal station "Dumont d'Urville", through a two weeks over-snow traverse (protecting the equipment from jolts, vibrations, thermal shocks and low temperatures);
- the low availability of operators inside the base and total isolation for about 8-9 mounts during wintertime.

## 3. AMICA Equipment

AMICA is a multiband camera provided of two detectors. A InSb 256x256 CRS-463 Rayteon array and a Si:As 128x128 DRS Tech. array are used for imaging in the NIR (2-5 μm) and MIR (7-28 μm) regions, with a set of KLMNQ filters. A complete equipment provides all the necessary support to the camera operation. To perform safe and unattended observations in such peculiar conditions, it is necessary to meet all environmental, cryogenic and functional requirements, paying attention to failures recovery and maintenance. After solving a large number of issues, the development of the control system is currently at its final stage. This is focused on the integration of the camera with all hardware and software subsystems and to the verification of the correct execution of automatic procedures that will be cooperatively carried out with IRAIT.

### 3.1. Hardware Layout

**Modularity and Subsystems**

The compact configuration allows the whole system to co-rotate with the telescope fork (Fig. 1). All the instrumentation, except for the front-end electronics (bias and clock filters, video signals pre-amplifier), is thermally controlled in separate insulated boxes. They have been designed to be easily mounted and transported by cranes used at Dome C. As shown in Fig. 2, the components distribution has been carefully studied (taking into account thermal and size constraints) in order to ensure as much as possible their easy access and a fast replacement in case of malfunction. A schematic description of the AMICA control system (hardware layout and data connection) is shown in Fig 3.

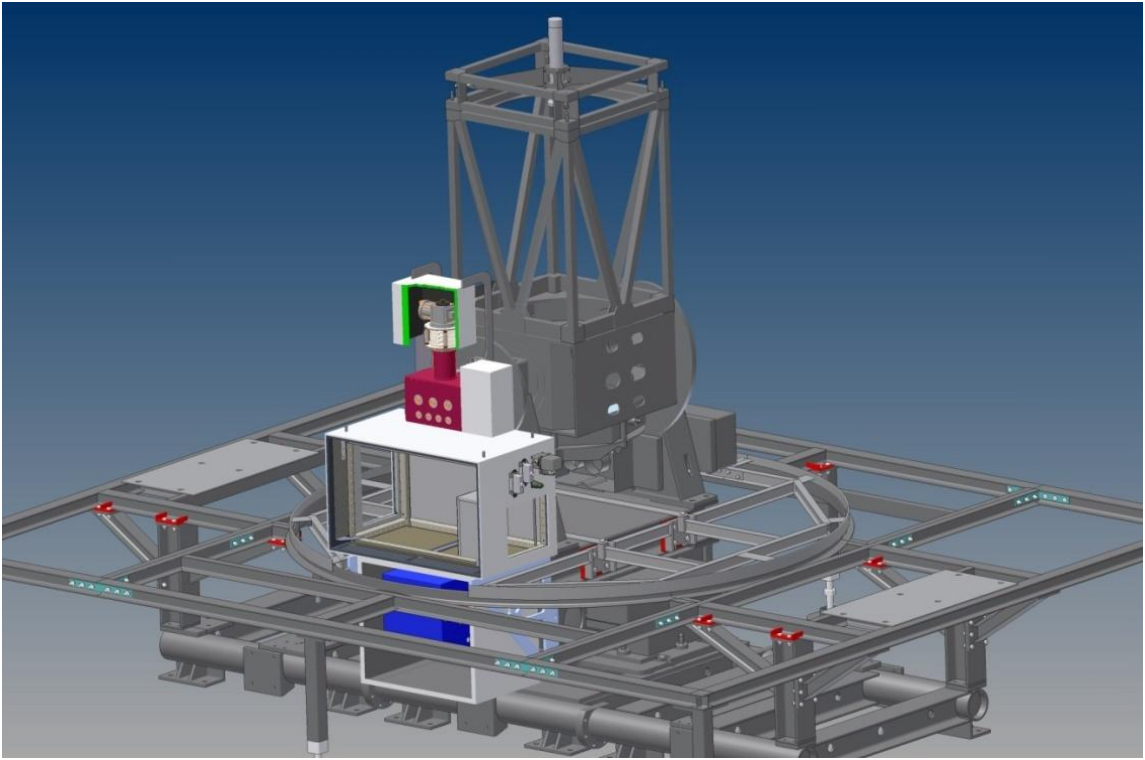

**Figure 1:** 3D rendering of the AMICA equipment mounted at the IRAIT fork. Most of the instrumentation is enclosed in thermal conditioned boxes (white) while the cryostat (red) is mounted at the telescope focus. Two small cabinets contain the cryo-cooler cold head and some vacuum system components. The box below encloses the LCU, the read-out electronics, controllers and camera auxiliary devices. The cryo-compressor (blue) is mounted below the rotating floor while the M2 wobbling system (light grey) is mounted at the IRAIT top ring.

The detectors control electronics (ACQ) has been developed cooperatively by INAF's departments (Padova, Teramo, Torino) and the Skytech Srl manufacturer [7]. It allows a pixel exposure time down to 0.7 μsec (16 bit ADC, up to 8 Mpx s$^{-1}$) with a minimum frame time of 2.9 msec for the MIR channel. It is composed of a digital programmable sequencer (PMC) installed in the cPCI Local Control Unit (LCU), optically connected through a 1.2 Gbaud fiber link with a separated rack. This hosts three further subunits: the SPC board provides the detectors clocks, while two DCS boards are used for the biases generation and for the correlated multisampling of the video signals.

Since mid-infrared observations require fast chopping techniques using the wobbling secondary mirror (M2) of IRAIT (with frequency of 2-10 Hz and a frame rate up to 300 fps for the AMICA+IRAIT configuration), the co-adding and the sky subtraction of raw frames are performed in real-time mode during exposures, directly triggered through TTL lines by the PMC or (for redundancy) through a TCP/IP command interface.

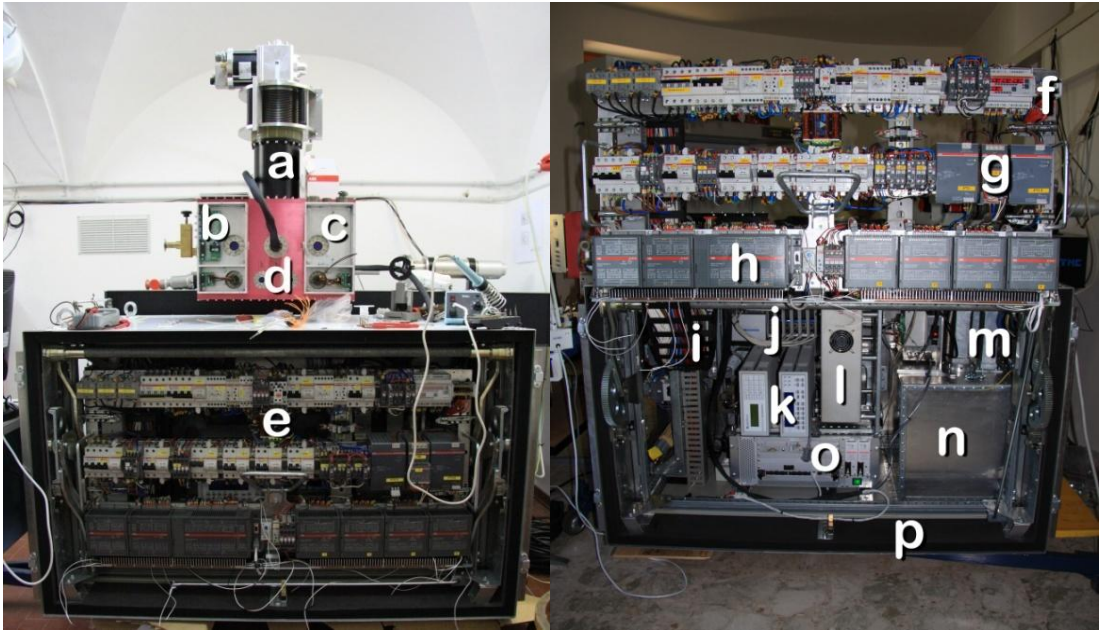

**Figure 2:** AMICA and the main thermal-conditioned box containing most of its equipment: a) cryocooler head; b,c) front-end electronics; d) cryostat; e) removable electrical panel (open on the right); f) multimeter; g) 24V redundant power supply; h) PLC with analog and digital I/O expansions; i) turbo-molecular vacuum pump, gauges and electromagnetic valve; j) Ethernet switch and MOXA converters; k) Lakeshore controllers; l) read-out electronics; m) heat dissipation system; n) rotary pump; o) LCU; p) insulating plastazote layer and alucor panels;

The IRAIT M2 subsystem has been manufactured by the Spanish NTE SA, which has also built a driver for the motion of the tertiary mirror (M3), in order to alternatively feed both the Nasmyth foci of the telescope . They have been tested to successfully operate down to -80°C, with the M2 driver able to perform fast and accurate pointing during chopping (up to 10 Hz) and off-axis imaging. The M2 driver in particular, has been fully integrated with the camera control system, in order to evaluate the overall accuracy and repeatability of the instrument, the settling time and the correct execution of automatic operations (e.g. observing modes configuration, focusing, etc.) during realistic simulations

A multi-level thermal control ensures the continuous monitoring and the conditioning of the instrumentation placed inside the boxes. In fact, besides the Environmental Control System (ECS, a high-level software application), a Programmable Logic Controller (PLC) is devoted to the management of a large number of analog and digital devices

(resistors, temperature and humidity probes, fans and contactors) to provide a low level active control of the conditions inside the cabinets. It can operate both autonomously (during outages of activity in which all other devices could be turned off) and cooperatively with the software running on the LCU. Moreover, the PLC is in charge of the boot and the shutdown of each component of the system (following well-defined safety procedures). Finally, a low-level passive electronics has been distributed inside the boxes to keep the minimum safe temperature in case of failure of all active systems.

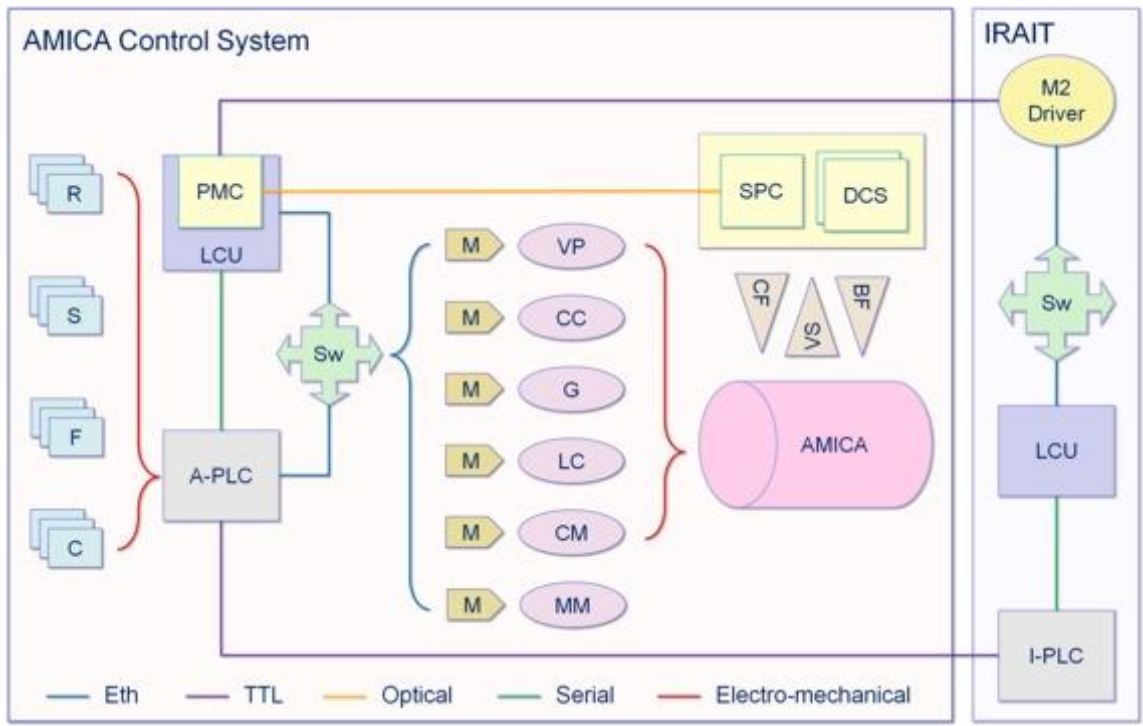

**Figure 3:** Hardware layout and data connections of the AMICA control system. The A-PLC manages resistors, sensors, fans and contactors (R,S,F,C) for the local environment control. A mutual heartbeat monitoring with the telescope PLC (I-PLC) is ensured by TTL lines. Auxiliary devices (vacuum pump, cryocooler, gauges, temperature controllers, cryomotors and multimeters – VP, CC, LC, CM, MM) are accessed by the LCU through serial/Ethernet MOXA converters. Clock and bias filters and video signal amplifiers (CF, BF, VS) are directly connect to the read-out electronics (SPC and DCS boards). This is linked (by the optical fiber) with the LCU in which a PMC board drives the M2 wobbling system through TTL lines. All the local control units and the Ethernet devices belong to the same LAN, extended by switches (Sw) and remotely connected to the base through cables and a redundant WiFi point-to-point link.

As discussed above, because of the high air rarefaction, not only the low temperatures, but also the low heat dissipation from electrical elements could seriously damage the instrumentation. For this reason, two cooling systems have been installed, composed of pipes passing through the boxes in which the forced ventilation (periodically alternated in direction) allows the internal environment to indirectly exchange the exceeding heat with the external one, avoiding thermal shocks (that could be induced by a direct exposure to the outside air) and the entrance of highly dangerous ice crystals.

## 3.2 Software Layout

The AMICA Control Software (ACSW) is an agent-based cooperative system, modeled under the principles of the OO Programming (C++, Java), using the graphical notation of the Unified Model Language (UML).

**Architecture**

The observatory operation is managed by the IRAIT OCS (Observatory Control System), which detain the observations scheduler and the weather control, retrieving information about all subsystems activity (e.g. busy status during acquisition, intensive backup, maintenance operations, etc.) as well as alarms due to dangerous events (low temperature inside the boxes, overheatings, malfunctions, etc.).
The ACSW architecture (Fig. 4) descends directly from the underlying modular hardware, with the aim to reduce the complexity level and to better identify all tasks that have to be assigned to each single- or multi-thread process. This distribution has led to the development of a multi-process system that allows to reduce the possibility of critical failures that could compromise the operation of all software modules (ACS, ECS, DCS, SCS and AAA, described below). Each subsystem is then controlled by one or more processes, which communicate through TCP/IP sockets, pipes or files.
The entry-point for the telescope scheduler is the Activity Control System (ACS), a server application that retrieves through a TCP/IP connection the required parameters to perform scheduled observations. It creates and dispatches macro-commands to the cooperative agents, managing and monitoring their activities and restarting them in case of unexpected hangs. Both telemetries and scientific data are periodically stored in a remote server inside the Concordia base, hosting a relational (MySQL) database and the image archive. Then, all data are preserved until they are shipped to European partners.
A real-time active control of the environment inside the insulated cabinets and the maintenance of the correct cryogenic conditions inside the cryostat are provided by the ECS. Since it operates cooperatively with the PLC (driving its activity), it can monitor the thermal, electrical and operational status of all the components of the AMICA hardware equipment. Moreover, camera auxiliary devices like heaters (for detectors thermal stability), temperature probes, cryomotors, the vacuum system and the cryocooler are indirectly managed by the ECS through their controllers (each of them being connected to the LAN by means of Serial/Eth MOXA converters). During the operation, the ECS collects all telemetry data, notices the ACS about the status of the system, giving the green light to the observations or asking to wait for the achievement of suitable operating conditions. Each dangerous event is therefore signaled to the ACS and in case of high safety risks, the ECS automatically takes the control of the system.

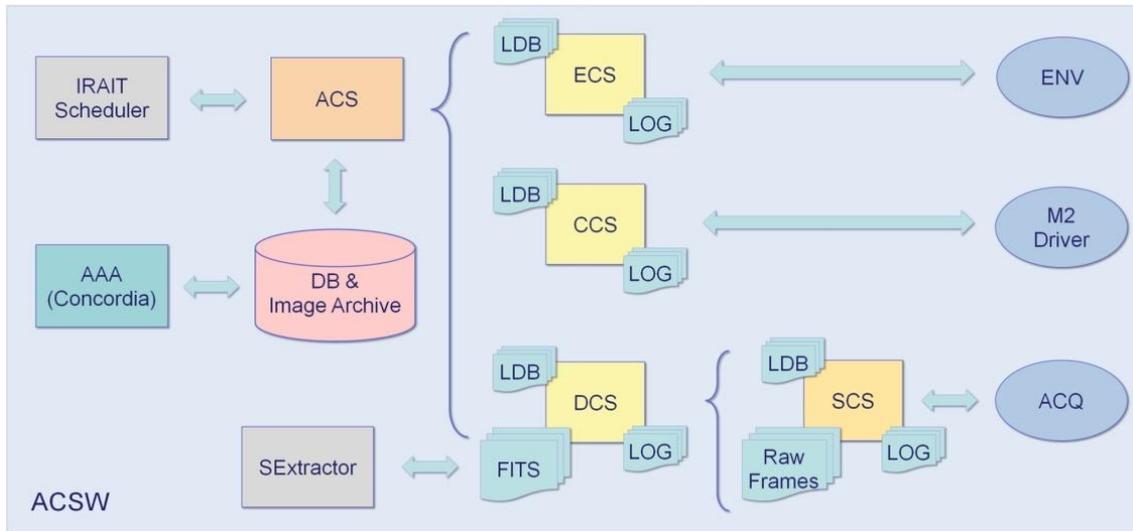

**Figure 4:** A deployment diagram showing the relationships among software and hardware modules in the run-time system. The ACS communicates with the IRAIT scheduler and manages four software modules (ECS, CCS, DCS, SCS) interfacing the hardware subsystems. A telemetry database and an image archive are maintained at the Concordia base, where a further application (AAA) monitors the whole system.

If it is not possible to recover its status, all unnecessary operations are stopped, turning off the corresponding devices and leading the environmental control to the PLC.
The configuration of the observing modes is operated by the Chopper Control System (CCS) that communicates with the IRAIT M2 subsystem through the LAN. It controls and verifies the correct position of M2 during acquisition (correcting for possible offsets) and supports the automated focusing procedures. It also monitors the subsystem activity gathering information on its thermal status and logging the ongoing operations.
The acquisitions are managed by the Detector Control System (DCS). It verifies the correct setting of the parameters (bias, clocks, $T_{exp}$, $N_{img}$, etc.), interfacing the read-out electronics through a further application (SCS) dedicated to the accumulation of the incoming raw frames (storing co-added and sky-subtracted images in a real-time process and filtering bad frames acquired during the motion of M2). In addition, raw data could be also saved for an off-line analysis of the M2 driver operation. Moreover, the DCS performs their descrambling, generating final FITS files, while a first pre-processing of the resulting images is initially obtained using the package SExtractor. It is used to estimate centroids, ellipticities and FWHMs of sources detected over the field and to obtain a preliminary pixels statistics. thus allowing the optimization of the observing parameters and of the focus quality, and the detection of bad images.

Finally, a remote Java web application (AMICA Activity Analyzer) is under development. Running on the control room workstation inside the base, it will provide statistics based on stored telemetry data for each component of the system, ensuring a full remote access to the instrumentation for local operators.

## 4. Aspects of the AMICA Control System

Generally, different levels of robotization can be distinguished for a system, depending on the capability to perform unattended operations, to resume its status after unexpected errors and to make use of some sort of intelligence to optimally operate for long-lasting periods (without human interaction). When all these properties coexist in such a system, it is usually referred to as "fully autonomous".

On the basis of these considerations, and taking into account the Antarctic conditions, a robotic observatory as that formed by IRAIT and AMICA, must necessarily be fully autonomous. Although the requirements on scheduler flexibility are less severe, thanks to the near-Pole condition and the possibility to observe throughout the year with a daily duty-cycle of 100% for wavelengths beyond 4μm (sec. 2.2), further issues have to be addressed for an "Antarctic" system.

### 4.1 Reliability and Test Activity

The robustness and reliability of the AMICA instrumentation has been the first point to be considered during its design. After a careful analysis of the environment properties and the possible risks that could arise from such a climate, a great attention has been paid to the choice of suitable devices available from the industrial automation market. In fact, the development of custom Antarctic-proof instrumentation would have had otherwise so strict requirements (more similar to those used in space engineering) that the total cost of the project would not been sustainable, thus losing all the advantages that had motivated the project itself. For this reason, all components have been selected in order to ensure the maximum resistance to the Antarctic conditions in terms of vibrations and shocks resistance, operating and storage ranges for temperature, pressure and humidity values. In addition, several thermal studies have been carried out about the materials and the thickness of the boxes insulating layer to minimize changes of their internal temperature. It has been considered suitable a thermal configuration that will maintain about +20 °C inside the cabinets with an external temperature of -60 °C

(winter average temperature), and with an average thermal input given by the dissipated heat from the operating electronics of about 750W. Thanks to this insulation, a low-level thermal control is able to maintain with a low power consumption, the internal temperature of the boxes between the storing range of values of the critical components (~ 0-50°). This estimation has been confirmed performing realistic simulations on the behavior of insulated, warmed boxes, through a climatic chamber built at INAF-Teramo (ANTARES – Antarctic Environment Simulator). It allows to reproduce climate conditions worse than those occurring at Dome C ($T_{min}$= -93.4 °C, P ~ 600 mbar, RH ~ 5%) in a ~200 ℓ volume. Thanks to these simulations, several effects have been studied as for example the wind chill effect induced by fans activation, the bias voltages drift with temperature, the failure of Buna O-ring vacuum seals below -43°C, etc.

Constraints on the total available space and instrumentation accessibility have led to the design of a very compact, complete and modular system, paying attention to the logistic facilities (e.g total power consumption, mounting and dismounting procedures).

Moreover, redundancy of critical components has been applied when possible, in particular for power supplies and electrical and data connections (TTLs, Ethernet, etc.), to continue operating even in case of damaging of any system element.

Despite the great attention that has been paid to the achievement of a highly reliable system, experiences in past Antarctic campaigns have highlighted the importance to provide each instrumentation with spares of sensible components. For this reason most of the elements constituting the AMICA equipment have been duplicated.

Multiple levels of control will ensure the instrumentation safety. Hardware devices and software systems will cooperate to achieve and maintain suitable conditions inside the AMICA conditioned boxes. In addition, all activities will be further monitored both through the LAN and the mutual exchange of "heartbeat" signals among programmable devices (i.e. LCUs and PLCs) belonging to the camera and the telescope subsystems.

Finally, several long tests on the control software have been performed, with the aim to prevent Single Points of Failure (SPOF), thus allowing the recovery of possible malfunctions and successfully executing all system tasks (acquisition and chopping management, image pre-processing, data storing, environment conditioning, remote monitoring, etc.).

**4.2 Future work**

The development of the whole control system is at its final stage. All hardware subsystem have been separately tested while the preliminary release of the control software allows to automatically execute (simulated) scheduled observations. Next steps will consist of the integration of the ECS module and the verification of the reliability of the thermal control, cooling the insulated boxes down to Antarctic temperatures.

Improvements in the DCS are required for the accomplishment of the pre-processing pipeline while further issues will be addressed to optimize the interaction between IRAIT and AMICA. Finally, the Activity Analyzer will be endowed with graphs and diagrams showing real-time information on the system activity and providing statistics that will be used to detect deviations from the expected behavior and to prevent malfunctions.

## 5. Conclusions

The excellent atmospheric properties of Dome C for infrared astronomy, allow to achieve better observing performances than any observing temperate site. Despite these exceptional advantages, several difficulties arise from its extreme environment. The AMICA project takes up the great challenge of the development of a highly reliable Antarctic instrumentation. For this reason, suitable and innovative solutions provide the necessary conditions to its robotization. After the conclusion of the integration stage and further test activities, the camera and its equipment will be shipped to Dome C for the accomplishment of the fully autonomous observatory.

## References

[1] G. Tosti, G. Nucciarelli, M. Bagaglia, et al., "The International Robotic Infrared Telescope (IRAIT)", *Proceedings of the SPIE*, Volume 6267, pp. 62671H, 2006.
[2] O. Straniero, M. Dolci, A. Valentini, et al. "AMICA: The First camera for Near- and Mid-Infrared Astronomical Imaging at Dome C", *EAS Publications Series*, Volume 25, pp.215-220, 2007.
[3] J.S. Lawrence, "Infrared and Submillimeter Atmospheric Characteristics of High Antarctic Plateau Sites", *The Publications of the Astronomical Society of the Pacific*, Volume 116, Issue 819, pp. 482-492, 2004.
[4] V.P Walden, M.S. Town, B. Halter, J.W.V. Storey, "First Measurements of the Infrared Sky Brightness at Dome C, Antarctica", *The Publications of the Astronomical Society of the Pacific*, Volume 117, Issue 829, pp. 300-308
[5] Burton, M.G., Lawrence, J.S., Ashley, M.C.B., et al.," Science programs for a 2-m class telescope at Dome C, Antarctica: PILOT, the Pathfinder for an International Large Optical Telescope", *Publications of the Astronomical Society of Australia*, Volume 22, Issue 3, pp. 199-235, 2005.
[6] R. Briguglio, G. Tosti, M. Busso, et al., "Small IRAIT, Telescope operations during the polar night", *Proceedings of the SPIE*, Volume 7016, pp. 70160H-70160H-12, 2008.
[7] Bortoletto, D. Magrin, C. Bonoli, et al., "Control System for the AMICA infrared camera", *Proceedings of the SPIE*, Volume 7014, pp. 70143A-70143A-10, 2008.